\newcommand{\edition}{ed01}
\newcommand{\version}{arXiv:\ v4,\ \ \today\ (\edition)}

\documentclass[10pt,
onecolumn,%
oneside,%
floats,%
aps,%
prd,%
nobibnotes,%
nofootinbib,%
amsmath,%
amssymb,%
amsfonts,%
superscriptaddress,%
]{revtex4}

\usepackage[utf8]{inputenc}
\usepackage{graphicx,array,dcolumn}
\usepackage{cases}
\usepackage[paperwidth=210mm,paperheight=297mm,centering,hmargin=1.8cm,vmargin=2.5cm]{geometry}
\usepackage{enumerate}

\usepackage{hyperref}

\usepackage[normalem]{ulem}
\usepackage{soul}
\usepackage{bm}
\usepackage{bold-extra}

 \def\la{\mathrel{\mathpalette\fun <}}

\def\fun#1#2{\lower3.6pt\vbox{\baselineskip0pt\lineskip.9pt
\ialign{$\mathsurround=0pt#1\hfil ##\hfil$\crcr#2\crcr\sim\crcr}}}

\newcommand{{\SD}}{\rm SD}

\newcommand{{\Mc}}{\mathcal{M}}

\newcommand{\vesig}{\mbox{\boldmath${\rm \sigma}$}}

\newcommand{\veV}{\mbox{\boldmath${\rm V}$}}

\newcommand{\veR}{\mbox{\boldmath${\rm R}$}}

\newcommand{\lan}{\langle}
\newcommand{\ran}{\rangle}

\newcommand{\lb}{\left(}
\newcommand{\rb}{\right)}

\def\-g{\sqrt{-g}}

\newcommand{\be}{\begin{equation}}
\newcommand{\ee}{\end{equation}}

\newcommand{\ben}{\begin{equation*}}
\newcommand{\een}{\end{equation*}}

\newcommand{\bea}{\begin{eqnarray}}
\newcommand{\eea}{\end{eqnarray}}

\renewcommand\rho{\varrho}

\begin{document}



\title{\sc \Large{Lindblad and Bloch equations for conversion of a neutron into an antineutron}}

\thanks{\version} 


\author{\firstname{B.O.}~\surname{Kerbikov}\medskip}

\email{borisk@itep.ru}

\affiliation{Alikhanov Institute for Theoretical and Experimental Physics,
Moscow 117218, Russia \smallskip}

\affiliation{Lebedev Physical Institute, Moscow 119991, Russia \smallskip}

\affiliation{Moscow Institute of Physics and Technology, Dolgoprudny 141700,
Moscow Region, Russia \bigskip}

\date{\today}

\begin{abstract}
\noindent We propose a new approach based on the Lindblad and Bloch equations for the density matrix to the problem of a neutron into an  antineutron conversion. We consider three strategies to search for conversion: experiments with trapped neutrons, oscillations in nuclei, and quasi-free propagation. 
We draw a distinction between $n\bar n$ oscillations in which the probability that a neutron transforms into an antineutron depends on time according to the sine-square law and the non-oscillatory overdamped $n\bar n$ conversion. We show that in all three cases  decoherence due to the interaction with the environment leads to non-oscillatory evolution.
  \bigskip 

 
\end{abstract}

\begin{flushright}
INT-PUB-18-015
\end{flushright}

\maketitle

\large

\section{Introduction \label{intro} }

The observation of neutron transformation to antineutron would be a  discovery
of fundamental importance which would reveal the existence of  physics beyond the SM. The possibility of  $n-\bar n$ transformation was suggested almost half a century ago \cite{1a,2a,3a}.  This process was extensively discussed in a recent
paper \cite{01} in theoretical and experimental aspects. There are three
possible experimental settings aimed at the observation of neutron-antineutron
oscillations. The first one is to use the neutron beam from a reactor or from a
spallation source.  The beam propagates a long distance to a target in which a
possible $\bar n$ component annihilates and  thus being detected. The second
one is to  establish a limit on nuclear stability because $\bar n$ produced
inside a  nucleus will blow it up. The third option is to use ultracold
neutrons (UCN) confined in a trap. The first method was used in the experiment
performed in the early 1990's at the Institute Laue-Langevin (ILL) in Grenoble.
It established the current best limit of $\tau_{n-\bar n} > 0.86\cdot 10^8$ s
\cite{02}. The internuclear searches for $n$-$\bar n$ oscillations were performed
by several experimental groups  \cite{03,7a,8a}.
The lower limit on the bound neutron lifetime obtained in these experiments may under certain model assumptions be recasted into the equivalent free neutron life times equal to $1.3\cdot 10^8 s$ \cite{03}, $2.7 \cdot 10^8 s $\cite{7a}, and 1.23$\cdot 10^8 s$ \cite{8a}.

All three types of $n-\bar n$ oscillation search experiments require a clear
and unambiguous  theoretical description within a coherent formalism.

The problems which arise in data analysis include:
\begin{description}
    \item[a)] the role of the interaction with the trap walls for UCN confined
    in a trap,
    \item[b)]  the  renormalization of the  free-space oscillation time due to
    nuclear environment,
    \item[c)]the effects caused by collisions with atoms and molecules of the
    residual gas in free and bottled neutron experiments.
\end{description}

Problems \textbf{a)} and \textbf{b)} were investigated by many authors (see \cite{01} and references below) and may be considered to be solved at least in the first approximation. The situation with the problem \textbf{c)} is much more unfavorable. The first idea which comes to mind is to use the Fermi potential. However this approach has a fundamental flaw in application to the low density gas. In experiments with thermal, and even with UCN neutrons, the neutron wavelength is much smaller than the average intermolecular distance. The neutron ``can see'' the individual gas molecules which rules out the use of Fermi potential.

The aim of the present work is to develop a formalism which gives a correct solution to problem \textbf{c)}, and more than that, allows to treat the three problems \textbf{a)}, \textbf{b)}, and \textbf{c)} by essentially the same equations. Our approach is based on the density matrix formalism \cite{04,05,06}. The problem of oscillations in a two-state physical system interacting with the environment was for the first time formulated and solved using the density matrix in a seminal paper by G.~Feinberg and S.~Weinberg \cite{07} devoted to muonium to antimuonium conversion. Since then the density matrix has been used to describe oscillations in a wide range of physical systems \cite{08,09,10,11,12,13,14, 20a,21a,22a,23a}. The first steps in the analysis of the neutron to antineutron transitions within this framework was done in \cite{15}.   
 
To find  the oscillation probability one has to derive the time dependence of $|\Psi_n(t)|^2$ and $|\Psi_{\bar{n}}(t)|^2$. The character of the time evolution is determined by the interplay of two parameters --- the transition matrix element $\varepsilon = \tau^{-1}_{{n}\bar{n}}$ and the damping parameter $\lambda$ which is the measure of the wave function ``reduction'' by the environment, -- see below. In free space $\lambda=0$ and the time evolution of $|\Psi_{\bar{n}}(t)|^2$ is sine squared oscillations (ignoring $\beta$-decay). This is a nonphysical picture since in any experiment the neutron-antineutron system  is immersed in  the environment. This may be trap walls, nuclear matter, or the residual gas. For any conceivable experiment oscillations die away due to decoherence. This is almost obvious for transitions in nuclei and in the trap. We show that oscillations are washed out in the residual gas with the lowest attainable density. The point is, that the mixing matrix element $\varepsilon=\tau^{-1}_{{n}\bar{n}}\lesssim10^{-23}$ eV is much smaller than the damping parameter induced by the environment. The non-oscillating equation for $|\Psi_{\bar{n}}(t)|^2$ valid at all times will be derived. Even in the gedanken experiment lasting for an infinite time without $\beta$-decay the system will never turn from the pure neutron to the pure antineutron state.

The paper is structured as follows. In Sec. II I introduce the density matrix formalism and Bloch equation. As a warming up exercise the derivation of the standard expression for the probability of the free $n-\bar{n}$ oscillations in vacuum is presented. In Sec. III the decoherence due to collisions with the trap walls in the experiments with UCN is discussed. The equation for the Bloch vector has the form of an equation  for the oscillator with friction. The friction parameter is inversely proportional to the time interval between collisions. In Sec. IV the Lindblad equation is introduced for the evolution of the reduced density matrix of an open system. Connection between the Bloch and the Linblad equation approach is clarified. The damping parameter is expressed in terms of the amplitudes of $n$ and $\bar{n}$ interaction with the environment particles and the collision frequency. In Sec. V the rate of neutron-antineutron transitions in nuclei is derived. In Sec. VI the decoherence due to interaction with the residual gas is investigated and it is shown that quantum damping destroys oscillations unless the residual gas pressure is unrealistically low. In Sec. VII the results are summarized. 

\section{Density matrix formalism. Oscillations in vacuum}

The density matrix formalism is a natural way to describe the quantum system in
a contact with the environment \cite{05,06,16,17}. Interaction in a medium breaks
the coherence  of the  propagation making the description in  terms of the wave
function impossible. The $n-\bar n$ system gets entangled with the environment. In order to set the scene  for the discussion of points
{\bf (a), (b)} and {\bf (c)}, consider first $n-\bar{n}$ oscillations in vacuum.
Needless to note that the well-known equation derived below may be obtained
with much less efforts using the Schrodinger equation.

In vacuum the state of a system is a pure one with the wave function \be | \psi
\ran = \varphi_1 (t) |n\ran + \varphi_2 (t)| n'\ran. \label{1}\ee Later we
shall identify the states $|n\ran$ and $|n'\ran$ with either neutron and
antineutron, or neutron and mirror neutron. The density operator reads \be
|\psi \ran \lan \psi| = \varphi_1\varphi_1^* | n\ran \lan n| +
\varphi_1\varphi_2^*| n\ran \lan n'| + \varphi_2\varphi_1^*| n'\ran \lan n| +
\varphi_2\varphi_2^*| n'\ran \lan n'| .\label{2}\ee

In  the matrix form one can write \be \hat \rho (t) = \left( \begin{array} {ll}
  \varphi_1\varphi_1^* &
  \varphi_1\varphi_2^*\\\varphi_2\varphi_1^*&\varphi_2\varphi_2^*\end{array}\right)
  \equiv\left( \begin{array}{ll}
  \rho_{11}&\rho_{12}\\\rho_{21}&\rho_{22}\end{array}\right).\label{3}\ee

In vacuum and without decay the hamiltonian has the form \be \hat H = \left(
\begin{array}{cc} E+\Delta_1 &\varepsilon\\\varepsilon&E+\Delta_2
\end{array}\right),\label{4}\ee

where $\varepsilon $ is a mixing parameter. At this point we do not specify the
origin of the level-splitting parameters $\Delta_i, i=1,2$. The density matrix
satisfies the von Neumann-Liouville equation \cite{05,06}. \be i\,\frac{d\hat
\rho}{dt} = [\,\hat H, \hat\rho\,].\label{5}\ee This yields four coupled linear
differential equations for the four components of $\hat\rho$. The time evolution of $\hat\rho$ may be represented
in the vector form of the Bloch equation \cite{18}. The real Bloch 3-vector
$\veR$ is introduced by the expansion of the density matrix over the Pauli
matrices \be \hat\rho = \frac12\,(1+\veR\vesig),\label{6}\ee \be \veR = \left(
\begin{array}{c} \rho_{12} + \rho_{21}\\ -i\, (\rho_{21} - \rho_{12})\\
\rho_{11} - \rho_{22}\end{array}\right).\label{7}\ee

The von Neumann-Liouville equation (\ref{5}) is equivalent to the following
equation of motion for the Bloch vector \be \dot{\veR} = \veV\times
\veR,\label{8}\ee where \be \veV = \left(\begin{array}{c}
2\varepsilon\\0\\d\end{array} \right), \label{9}\ee here $d=\Delta_1-\Delta_2$ is the level splitting caused by the ambient magnetic field.
Equation (\ref{8}) describes the precession of the Bloch vector $\veR$ around
the ``magnetic field'' \veV. According to this equation the length of $\veR$
does not change.  In particular, this means the absence of decoherence.
Decoherence is a process in which the off-diagonal elements of the density
matrix are reduced. Decoherence does not happen as soon  as the system is
isolated from the environment. In the component form the Bloch
equation reads \be \dot{R}_x =-dR_y, ~~ \dot{R}_y = dR_x -2\varepsilon R_z, ~~
\dot R_z=2\varepsilon R_y.\label{10}\ee

We shall solve these equations with the initial condition $\veR (t=0) =
(0,0,1)$ which means that the system is initially in the $|n\ran $ state.
Taking the derivative of the equation (\ref{10}) for $R_y$, one gets

\be \stackrel{\cdot\cdot}{R}_y=-\Lambda^2R_y,~~ \Lambda^2
=d^2+4\varepsilon^2.\label{11}\ee

The solution of (\ref{11}) with the above initial condition is \be R_y (t) =
R_y (0) \cos \Lambda t + \Lambda^{-1} [dR_x(0) - 2 \varepsilon R_z (0)] \sin
\Lambda t. \label{12}\ee Substitution into (\ref{10}) for $R_z(t)$ and keeping
in mind the initial conditions, one obtains \be R_z (t) = \rho_{11} -\rho_{22}
= 1- \frac{8 \varepsilon^2}{\Lambda^2 }\sin^2 \frac{\Lambda}{2} t.\label{13}\ee
From (\ref{13}) and the normalization condition $\rho_{11}+\rho_{22}=1$, we
obtain the well-known equation describing oscillations in vacuum \be \rho_{22}(t) = | \psi_{n'} (t)|^2 =
\frac{4\varepsilon^2}{\Lambda^2} \sin^2 \frac{\Lambda}{2} t. \label{14}\ee

In view of a tiny value of the mixing parameter $\varepsilon = \tau^{-1}_{n\bar
n}\la 10^{-23} $ eV \cite{01}, the level splitting  $d$ caused by the ambient
magnetic field suppresses oscillations for any conceivable experiment. In an
ultra-short time limit $\Lambda t/2 \ll1$ (\ref{14}) yields \be \rho_{22} \simeq
\varepsilon^2 t^2.\label{15}\ee

The time evolution law (\ref{15}) does not allow to define the transition
probability per unit time  \cite{05}.

  The von Neumann-Liouville equation (\ref{5}) with a
Hermitian hamiltonian describes the unitary evolution of the density matrix.
With the account of $\beta$-decay this property gets lost. This amounts to a
factor  $\exp (-t/\tau_\beta)$ in front of (\ref{14}). All four elements of
$\hat \rho$ are exponentially decaying with equal rate. The  Bloch vector
$\veR$ shrinks in  length but the oscillation pattern remains intact. 

It is important to  make  a distinction between  the suppression of oscillations due to  level   
splitting  and the decoherence which is also called the quantum damping \cite{17}. The key  point of 
decoherence is  that the off-diagonal elements of the
density matrix  are damped if the interaction within the  environment is different for  $|n\ran $ and  $|\bar n'\ran$ states  \cite{28a}. This is obviously the case for $n-\bar n$ system. In the next section  we consider the simplest example -- the  interaction of neutrons with the trap walls.  
\

\section{Decoherence due to collisions with the  trap walls}

Before we start to consider possible decoherence due to the interaction with
the trap walls a general remark is in order. Decoherence drastically suppresses
oscillations if the collision rate with the environment in much higher the
oscillation frequency $\varepsilon$. A less stringent condition is that at
least one collision should take place to alter the oscillation pattern. Below
we shall formulate these conditions in a clear-cut form for each situation
under consideration.

The process of $n-\bar n$ oscillations for UCN (ultracold neutrons) trapped in
a storage vessel has been studied in a number of papers, e.g.,
\cite{19,20,21,22,23,24,25,26}. Here we want to look at this process from a new
angle. The trap walls take the role of the
eivironment. The simple model of a trap is a one-dimensional square
well. Let $\tau_i$ be the  time interval between the $(i-1)$-th and $i$-th
reflections. We also introduce the average time between collisions $\tau=t/n$,
$n$ is the number of collisions and we assume that $n\gg 1$. This is not a
necessary condition for the decoherence but it allows to obtain a solution in a
closed analytic form. We also note that in the wave-packet formalism one has to
introduce somewhat different approach \cite{27}. It will be assumed that the
collision with the wall is instant and that the antineutron component is
absorbed at the wall without reflection. These two conditions have been
loosened in \cite{24}. We solve the problem making use of the Bloch equations
(\ref{10}).  Returning to (\ref{1}) we identify the state $|n\ran$ with neutron
and $|n'\ran$ with  antineutron. The splitting $d$ between the two states is
for simplicity discarded. One more notation is needed. Even though the
interaction with the trap wall is assumed to proceed instantaneously, one
should distinguish the time $\tau_i$- just before the $i$-th  collisions and
$\tau_{i+}$ just after. At $t=0$ the system starts to evolve from the state
$|n\ran $ = neutron. According to (\ref{10}) at $0<t<\tau_1$ -- the evolution
proceeds according to \be \dot{R}_z = 2 \varepsilon R_y, ~~ \dot{R}_y = - 2 \varepsilon
R_z.\label{16}\ee

It terms of the wave functions this corresponds to \be \psi_n(t) = (\psi_n(0)
\cos \varepsilon t - i \psi_{\bar n}(0) \sin \varepsilon t) \exp (- i E
t),\label{17}\ee

\be \psi_{\bar n}(t) = (- i \psi_n(0) \sin \varepsilon t +   \psi_{\bar n}(0)
\cos \varepsilon t) \exp (- i E t).\label{18}\ee

We remind that we set $d=0$ and $\tau_\beta =\infty$. At $t= \tau_{1-}$ the
solution of (\ref{10}) reads \be R_z(\tau_- )=\cos 2 \varepsilon\tau_{1-},~~
R_y(\tau_{1-}  )=-\sin 2 \varepsilon\tau_{1-}.\label{19}\ee After the first
collision at $t= \tau_{1+}$ one has \be R_z(\tau_{1+} )=\cos^2
\varepsilon\tau_{1+},~~ R_y(\tau_{1+}  )=0.\label{20}\ee

Note that for neutron-mirror-neutron oscillation \cite{27} $R_z(\tau_{1+})=R_z(\tau_{1-} )$ and the process proceeds differently. This  will be
  discussed elsewhere. The answer for  $   R_z(\tau_{1+}+ \tau _{2+} +... + \tau_{n+}
  )\equiv   R_z(\tau_{n+} )$ now seems evident
\be  R_z(\tau_{n+} ) = \prod^n_{k=1} \cos^2  \varepsilon\tau_{k+}.\label{21}\ee Note that $\varepsilon \simeq 10^{-8}
s^{-1}, \tau \simeq 0.1$s \cite{28} and therefore $\varepsilon \tau \ll 1$.
Averaging over time intervals between collisions, one obtains\footnote{The
author is grateful to M.I.~Krivoruchenko for the discussion of this point, see
also \cite{26}. An alternative averaging  equation is used~in~\cite{09}.}

\be R_z = \prod^n_{k=1} \int \frac{d\tau_k}{\tau} \exp \left( -
\frac{\tau_k}{\tau}\right) \cos^2 \varepsilon \tau_k = \left( \frac{ 1/\tau^2+
2 \varepsilon^2}{ 1/\tau^2+ 4 \varepsilon^2} \right)^{n} \simeq \exp
(-2\varepsilon^2\tau t).\label{22}\ee

The most important quantity is the admixture of $\bar n$ at $t= n\tau_{n-}$,
i.e., before the $n$-th collision. It is given by\be |\psi_{\bar n} (\tau_{1 }+ \tau _{2 } +... + \tau_{n-})|^2 \equiv | \psi_{\bar n}
   (\tau_{n-})|^2 = \sin^2\varepsilon \tau_{n-}  \prod^{n-1}_{k=1} \cos^2(  \varepsilon\tau_{k+})
   \simeq \varepsilon^2\tau^2_n \exp (-2 \varepsilon^2 \tau t), \label{23}\ee
where the number of collisions $n$ is assumed to be large and hence $(n-1) \tau
\simeq n\tau =t$, and $\varepsilon \tau \ll 1$. The annihilation probability
after $n$ collisions is equal to \be P_a (\bar n) =\varepsilon^2  \exp (-2
\varepsilon^2 \tau t) n \left(\frac{1}{n} \sum^n_{k=1} \tau^2_k\right)=
\varepsilon^2 \exp (-2 \varepsilon^2 \tau t)\frac{\lan \tau^2\ran}{\tau} t \simeq
\varepsilon^2\frac{\lan \tau^2\ran}{\tau} t\simeq \varepsilon^2\tau
t.\label{24}\ee The last result is a well-known one \cite{24}.

At this point one may question whether the factor $\exp(-2 \varepsilon^2 \tau
t)$ should  be kept in the above equation since $ \varepsilon^2 \tau t\sim
10^{-14}$ even of $t$ is taken  equal to the free neutron life time. To answer
this question we  return to (\ref{16}) and introduce an additional damping
parameter $\lambda$ into the equation for $\dot{R}_y$ \be \dot R_z =
2\varepsilon R_y, ~~\dot R_y = -2\varepsilon R_z-\lambda R_y.\label{25}\ee

The factor $\lambda$ should not be confused with the $\beta$-decay constant
$\gamma=1/\tau_\beta$ which enters into the equations for all three components
of the Bloch vector (we remind that $\beta$-decay is temporary discarded). From
(\ref{25}) one obtains the following equation for $R_z$

\be \stackrel{\cdot\cdot}{R}_z+\lambda\dot R_z+ 4 \varepsilon^2 R_z
=0.\label{26}\ee

This is an equation for the oscillator with friction. Assuming that at $t=0$
the system is in a state $|n_1\ran = n$, so that $\rho_{11} (t=0)=1, $
 $\rho_{22} (t=0)=0 $, the solution of (\ref{26}) may be written in the
 following form
 \be R_z (t) = e^{-\frac{\lambda}{2}t} \left[ \frac{\frac{\lambda}{2} + \Omega}{2 \Omega} e^{\Omega t}
 + \frac{-\frac{\lambda}{2} + \Omega}{2 \Omega} e^{-\Omega t}\right], \label{27}\ee
where $\Omega$ is given by \be \Omega^2 = \left( \frac{\lambda^2}{4}-4\varepsilon^2\right).\label{28}\ee

The solution (\ref{27}) corresponds to $\Omega^2 > 0$. Other regimes will be discussed in Section VI. For ``long'' times
$t \gg 1/\lambda$ the overdamping solution of (\ref{27}) is proportional to \be R_z \sim \exp
\left(-\frac{4\varepsilon^2}{\lambda} t\right).\label{29}\ee

The condition to match (\ref{22}) requires \be \lambda=
\frac{2}{\tau}.\label{30}\ee The typical time $\tau$ between collisions (time
of neutron free fight) is $\tau\simeq 0.1$ s  \cite{28}.

The following conclusions on decoherence may be deduced from the example of
oscillations in a trap.  Each interaction with the walls destroys the coherence
completely. The account of the final interaction time and partial
reflection of antineutrons can make this statement only partially true
\cite{24}. In terms of the Bloch vector the transverse components $R_x$ and
$R_y$ are nullified at each collision. The component $R_z$ satisfies equation
(\ref{26}) which is an equation of oscillation with friction.  From the
solution (\ref{27}) it follows that at any time \be \frac{|\psi_{\bar n} (t)
|^2}{|\psi_{  n} (t) |^2} \simeq \frac{4\varepsilon^2}{\lambda^2} \ll 1.
\label{31}\ee This is an overdamping regime previously discussed in \cite{29}
and \cite{07}. With $\lambda= 2/\tau$ (\ref{31}) yields \be \frac{|\psi_{\bar n}
(t) |^2}{|\psi_{  n} (t) |^2} \simeq  { \varepsilon^2}{\tau^2} \sim 10^{-18}.
\label{32}\ee which is in agreement with (\ref{24}). At first  glance this
makes the task to observe oscillations extremely difficult. The situation
becomes even worse if the energy splitting between neutron and antineutron due
to external magnetic field is taken into account. We also did not take into account decoherence due to the interaction with the residual gas  inside the trap. This subject will be considered in Sec. VI. Finally, we remind that the
above numerical estimates were obtained for a very simple model of the UCN
trap. As a last remark we rise a question of whether decoherence in hitting the trap walls may be avoided by a special choice of the wall material. As it was shown in \cite{24} the key parameter which leads to decoherence is the collision time, or time difference. The difference of time delay results in different phases of the reflection coefficients. In principle, it should be possible to find the materials which do not lead to decoherence (Valery Nesvizhevsky, private communication). 

In the two previous sections the evolution equations for the Bloch vector were written down without being guided by a rigorous mathematical formalism. Prior to considering the oscillations in nuclei and in the residual gas inside the experimental setup, it makes sense to give a glimpse at Lindblad equation which is a foundation of the Bloch vector evolution for an open system.

\section{Lindblad Equation for the Bloch Vector}

The time evolution of the effective density matrix of a subsystem interacting with the environment is described by the Lindblad \cite{30} (or Lindblad-Gorini-Kossakowaky-Sudarshan \cite{31}) equation. It has the following form (compare with Eq.(\ref{5}))

\be {\dot {\hat \rho}}(t) = \frac{d\hat
\rho(t)}{dt} = -i[\hat H, \hat\rho(t)]+\sum\limits_n\,[\,L_n\,\hat\rho(t)\,L_n^{\dagger}-\frac{1}{2}\,L_n^{\dagger}L_n\,\hat\rho(t)-\frac{1}{2}\,\hat\rho(t)L_n^{\dagger}L_n\,].\label{33}\ee
where $L_i$ are Lindblad operators which satisfy certain conditions \cite{32} but are not known a priori. A pedagogical derivation of the Lindblad equation may be found in \cite{32}. The extra terms in Eq.(\ref{33}) is a price to pay for the use of the reduced density matrix describing only the evolution of the subsystem density matrix. 

Lindblad equation  (\ref{33}) 
is a generalization for an open system of the evolution equation (\ref{5}), which in its turn may be written in the form (\ref{8}) in terms of the Bloch vector. It is natural to ask what is the equation for the Bloch vector corresponding to (\ref{33}).
The choice of the Linblad operators depends on the system dynamics, i.e., on the form of the Hamiltonian. For  an isolated $n\bar n$ system without $\beta-$decay the Hamiltonian is given by Eq.(\ref{4}). To include the interaction with the environment (nuclear matter, or ambient gas) we introduce the following quantities: $n$ -- the environment number density, $v$ and $k$ -- the relative velocity and momenta of the neutron and the medium particles, $f_1(\theta)$ and $f_2(\theta)$ -- neutron and antineutron elastic scattering amplitudes off the surrounding particles (nucleons, or gas molecules). For   UCN and thermal neutrons it is  reasonable to assume that the interaction is saturated  in the $s$-wave.   
The Linblad equation (\ref{33}) may be  rewritten in the following form \cite{33}

\be {\dot {\hat \rho}}(t) = -i (H\hat \rho-\hat \rho H^+) + L\hat\rho L^+ - \frac12 \{L^+L, \hat\rho\}, \label{34a}\ee
where $\{...\}$ is an anticommutator, and 
\be \hat H =  \left( \begin{array} {cc} E+\Delta_1 - \frac{2\pi}{k} nv Re f_1(0) - i \frac{\gamma}{2}& \varepsilon\\
\varepsilon& E+\Delta_2 - \frac{2\pi}{k} nv Re f_2(0) - i \frac{\gamma}{2}\end{array} \right).\label{35a}\ee
Here $\gamma$ is the  $\beta$-decay constant. Note that in \cite{07} the Hamiltonian contains $\frac{2\pi}{k} nv f_i (0)(i=1,2)$ without Lindblad terms in the evolution equation. The  representation (\ref{35a}) allows to write the equation in the Lindblad form. Both parametrizations lead to the same physical results.  The terms $\frac{2\pi}{k} nv Ref_i(0), ~~ (i=1,2)$ in (\ref{35a}) correspond to the energy shift related to  forward scattering, i.e., the real part of the index of refraction. The Lindblad operator is defined as \be L=\sqrt{nv}\,F, ~~ F=\left( \begin{array}{cc} f_1(\theta)&0\\0&f_2(\theta)\end{array}\right).\label{36a}\ee

Note that the interaction is assumed to proceed in the $s$-wave.

Next we introduce total, elastic and reaction cross sections $\sigma_{it}, $ $\sigma_{ie}, \sigma_{ir}$ for  neutron $(i=1)$ and antineutron $(i=2)$ interaction with the medium particles. With the definition of the Lindblad operator (\ref{36a}) and using the optical theorem $\sigma_{it} = \frac{4\pi}{k} Im f_i (0)$ we can transform the Lindblad equation (\ref{34a}) into a set of the  following   equations for the density matrix components
\be 
\dot{\rho}_{11} =- i\varepsilon (\rho_{21}-\rho_{12}) - (nv\sigma_{1r} +\gamma) \rho_{11}, \label{37a}\ee
\be 
\dot{\rho}_{12} =  i\varepsilon (\rho_{11}-\rho_{22}) - i (d+K)  \rho_{12} -(M+\gamma) \rho_{12}, \label{38a}\ee
\be 
\dot{\rho}_{21} =- i\varepsilon (\rho_{11}-\rho_{22}) + i (d+K)  \rho_{21} -(M+\gamma) \rho_{21}, \label{39a}\ee

\be 
\dot{\rho}_{22} =  i\varepsilon (\rho_{21}-\rho_{12}) - (nv\sigma_{2r} +\gamma) \rho_{22}, \label{40a}\ee

where \be K=-\frac{2\pi}{k} nv {\rm Re} (f_1(0)- f_2(0)) - 4\pi nv \operatorname{Im} f_1 f_2^*.\label{41a}\ee

\be M=   4\, \pi n v \left[\operatorname{Im} \left( \frac{f_1(0)+f_2(0)}{2k} \right) -  \operatorname{Re} \left(f_1 f_2^*\right)\right] \label{42a}\ee
and $d=\Delta_1-\Delta_2$. The system of equations (\ref{37a})-(\ref{40a})  with the quantities $K$ and $M$ defined by (\ref{41a}), (\ref{42a}) provides a complete description of the  neutron-antineutron system time  evolution. Without further simplifications, e.g., dropping certain terms, this system of equations can not be solved analytically. The first physically reasonable approximation is to neglect the inelastic interactions of the neutron, i.e., to put $\sigma_{1r}=0$. The inelastic channel for the antineutron may be identified with annihilation, $nv\sigma_{2r}=nv\sigma_a\equiv \lambda/2$ with the reason for the last notation to become clear shortly. Less obvious step is to discard the term $(nv\,\sigma_{2r})\rho_{22}=(nv\,\sigma_{a})\rho_{22}$ in equation (\ref{40a}). Here we follow the arguments presented in \cite{33a}, namely that this term corresponds to the rate of disappearance of $\bar{n}$ when its coupling $\varepsilon$ to the neutron is neglected. Then the system of equations (\ref{37a})-(\ref{40a}) may be replaced by the following equation of motion for the Bloch vector $\veR$ defined by (\ref{7})

\be \dot{\veR} =\left( \begin{array}{ccc}-(M+\gamma)& -(d+K) &0\\ (d+K)&-(M+\gamma)&
-2\varepsilon\\0& 2\varepsilon&-\gamma
 \end{array}\right)~\left( \begin{array}{c} R_x\\R_y\\R_z
\end{array}\right).\label{43x}\ee

Equation (\ref{43x}) may be written in the form first proposed by L.Stodolsky \cite{17}

\be \dot{\veR} = \veV \times \veR - D_T\,\veR_T - \gamma\,\veR, \label{44x}\ee

where 

\be \veV = \left( \begin{array}{c} 2\varepsilon\\0
\\d+K\end{array}\right)\!, ~~
D_T= \left( \begin{array}{cc} M &0\\0 & M
\end{array}\right)\!, ~~\veR_T = \left( \begin{array}{c}R_x\\R_y
 \end{array}\right)\!, ~~\gamma=\dfrac{1}{\tau_{\beta}}.\label{45x}\ee
 
According to (\ref{44x}) the medium induces an additional contribution into $V_z$ which corresponds to a supplementary ``magnetic field'' along the $z$-axis. In other words, $K$ corresponds to the energy shift due to the refraction index. The damping parameter $D_T$ characterizes the rate at which the classical environment destroys the off-diagonal elements of $\rho$, thus leading to a loss of coherence, or to a collapse of the density matrix. Due to the second and the third terms in (\ref{44x}) the Bloch vector $\veR$ shrinks in length. The quantites $K$ and $M$ defined by (\ref{41a}) and (\ref{42a}) may be following \cite{17} expressed as a real and an imaginary parts of the expression 

\be \Sigma
= nv\,\dfrac{i \pi}{k^2}\,(1-S^*_1S_2),\label{46x}\ee

\be K=\operatorname{Re}\Sigma,\qquad M=\operatorname{Im}\Sigma,\label{47x}\ee

where $n$ is the environment
number density, $v$ and $k$ are the relative velocity and momenta of the
neutron and the medium particles,  $S_i\,\,(i=1,\,2)$ are the scattering matrices in
the channels $|n\ran $ and  $|\bar n\ran $. 

Without further simplifications equation (\ref{43x}) for the Bloch vector can hardly be solved analytically \cite{42}. In particular, to obtain (\ref{26}) from (\ref{43x})-(\ref{44x}), one has to drop the term $(d+K)$ and $\gamma$. Taking the second time derivative of $R_z$ and denoting $\lambda=M=\operatorname{Im}\Sigma$, one immediately arrives at 

\be \stackrel{\cdot\cdot}{R}_z+\lambda\dot{R}_z+4\,\varepsilon^2\,R_z=0.\label{48x}\ee

Starting from the Lindblad equation (\ref{33}), defining the Lindblad operators according to (\ref{36a}), we have reproduced (\ref{26}). That is, we derived the basic equation (\ref{18}) of \cite{17}. The important point is that the ``friction'', or damping parameter $\lambda=M$ is by (\ref{42a}) expressed through the collision frequency and the amplitudes $f_1$ and $f_2$ of neutron and antineutron interaction with the environment constituents. Another point to mention is that ``friction'' is due to decoherence, or entropy \cite{20a,17}.

Equation (\ref{48x}) exhibits three different regimes depending on the relative magnitudes of the damping parameter $\lambda$ and the mixing parameter $\varepsilon=\tau^{-1}_{n\bar{n}}$. The character of the solution of (\ref{48x}) is determined by the sign of the quantity \be \Omega^2=\frac{\lambda^2}{4}-4\varepsilon^2. \label{49x}\ee If $\Omega^2 > 0$ (overdamping regime) interaction with the environment destroys the off-diagonal elements of the density matrix and the solution is given by (\ref{27}). The opposite (underdamped) case $\Omega^2 < 0$ corresponds to slightly or strongly damped oscillations depending on the relative magnitudes of ${\lambda}$ and $\varepsilon$. The case $\Omega^2 = 0$ (critical damping) implies a fine tuning of ${\lambda}$ and $\varepsilon$. Finally, for $\lambda = 0$ the system is in the free oscillation regime (\ref{14}).   

\section{Neutron-Antineutron Oscillations in Nuclei}

It is well known that in nuclei neutron-antineutron oscillations are strongly
suppressed (see \cite{01} and references therein). We shall show that the rate
of oscillation in nucleus and the neutron rate of disappearance can be obtained
from the equation (\ref{48x}). It will  lead us to the results obtained in \cite{29,34}
using the optical model and time-dependent coupled Schrodinger equation. The
present approach allows to clarify the physical picture behind these results.

Equation (\ref{48x}) is a truncated version of the complete matrix system
(\ref{43x}). Now comes another simplification. We consider the $\bar n$
annihilation in the nuclear medium as a dominant process and neglect much
weaker neutron interaction.

Both approximations can be easily lifted solving (\ref{43x}) numerically and
using for  $f_1$ and $f_2$ parametrization available in the literature.

Considering $s$-wave annihilation as the main source of decoherence one can write
\be
\lambda\simeq 4\pi\,n\,v\,\operatorname{Im}\dfrac{f_2(0)}{2k}=n\,v\,\sigma_a/2 \equiv \Gamma_a.
\label{50x}\ee
The factor $1/2$ in (\ref{50x}) reflects the fact that according to our assumption only one of the two components, namely antineutron, is active in the interaction with the environment. Note that the similar $1/2$ factor is present in \cite{07,09}. As an estimate one can take $v \sigma_a \simeq 50$ mb \cite{36}, $n \simeq 0.17 \text{ fm}^{-3}$, then $\Gamma_a \simeq 50 \text{ MeV}$ which agrees with the results of the detailed calculations \cite{35}. The damping parameter in nuclei $\Gamma_a \simeq 50$ MeV exceeds the mixing matrix element $\varepsilon \simeq 10^{-23}$ eV by thirty one orders of magnitudes. At $t \gg 1/\Gamma_a \simeq 10^{-23}$ s the term proportional to $\exp (-\Gamma_at/2)$ in the solution (\ref{27}) of the equation (\ref{48x}) can be dropped out and $R_z$ takes the form (see \cite{29}) \be
R_z(t)= \rho_{11}-\rho_{22} \simeq \exp\lb -\dfrac{4\varepsilon^2}{\Gamma_a}t \rb - \lb\dfrac{4\varepsilon^2}{\Gamma_a^2}\rb\exp \lb -\dfrac{4\varepsilon^2}{\Gamma_a}t \rb
\label{51x}\ee
The first term in (\ref{51x}) describes the time evolution of $|\Psi_n(t)|^2$. The disappearance lifetime is $T=\Gamma_a/4\varepsilon^2 \gtrsim 10^{32}$ yr. for $\varepsilon^{-1} \gtrsim 0.86 \cdot 10^8$ s from the ILL experiment \cite{02} which is in agreement with \cite{7a}. The antineutron component given by the second term in (\ref{51x}) is damped by a huge factor $4 \varepsilon^2/\Gamma_a^2 \sim 10^{-62}$ at all times (compare with (\ref{31})). As stressed in \cite{01} the above value of $T$ is of the same order of magnitude as the present sensitivity of experiments conducted in large underground detectors. 

\section{Neutron-Antineutron Transitions in Gases}

The existing direct limit on the oscillation time $\tau_{n-\bar{n}} > 0.86 \cdot 10^8$ s was obtained long ago in the experiment performed at ILL in Grenoble \cite{02}. Cold neutron beam with the average velocity $\sim 600$ m/s crossed a $76$ m vacuum tube in quasi-free conditions (see below) and in about $0.1$ s reached the annihilation target. The significantly improved experiment is under discussion for the realization in the European Spallation Source (ESS) \cite{01,35}. The role of the environment in experiments of this kind and in the experiments with trapped UCN is played by the low pressure residual gas. As it was explained in the Introduction the approach to the transitions in the residual gas based on the Fermi quasi-potential is inevitably flowered. The wavelength of the thermal neutron $\lambda \simeq 10^{-8}$ cm is four orders of magnitude smaller than the average intermolecular distance $n^{-\frac{1}{3}} \simeq 10^{-4}$ cm (for the ILL experiment). The density matrix treatment is a correct way to describe the neutron-antineutron transformations in gas.

According to our previous considerations decoherence arises when the two components (${n}$ and ${\bar{n}}$) interact with the medium differently \cite{28a}. The main source of decoherence is annihilation. The evolution of the Bloch vector component $R_z=(\rho_{11}-\rho_{22})$ is described by (\ref{48x}) with $\lambda$ given by the  equation (\ref{50x}). The quantity  $\lambda$ is the rate of inelastic (annihilation) collision frequency. Another source of decoherence is the difference between the neutron
and antineutron elastic scattering of gas. This effect is difficult to estimate without reliable information on the ${\bar{n}}$-gas scattering amplitude. Anyhow, annihilation is by far the leading source of decoherence. It is easy to see that the overdamping regime without oscillations is realized in any conceivable experiment on quasi-free propagation in gas. Indeed, the $\Omega^2 \leq 0$ condition (see (\ref{49x}))corresponds to \begin{equation} n \leq \frac{2 \varepsilon}{v \sigma_a} \simeq 10^6 \text{ cm}^{-3},\label{52x}\end{equation} which is more than $4$ orders of magnitude less than density of the residual gas in the ILL experiment \cite{02}. The result (\ref{52x}) was obtained taking into account the following considerations. The dominant fraction of the residual gas is molecular hydrogen. At room temperature the average thermal velocity of $H_2$ is $v_T \simeq 2 \cdot 10^5 \,\frac{\text{cm}}{s}.$ For thermal neutrons we can take the relative $nH_2$ velocity equal to $v \simeq 10^{-5}$ in natural units $c=1$. At low energy $v\, \sigma_a\,(\bar{n}\,p) \simeq 50 \text{ mb}$ \cite{36}. In view of (\ref{52x}) we also conclude that experiments performed with trapped UCN in recent years \cite{28,37,38} correspond to the overdamped regime. In the experiments of this kind there two sources of decoherence: interaction with the trap walls considered in Section III, and the collisions with the residual gas molecules. Necessary to note that the above experiments  were aimed at the observation of the neutron transformation to mirror neutron. The process is described by different equations \cite{27}. The lower limit of the corresponding oscillation time is $\tau_{n-n'} > 414 \text{ s}$ \cite{28}. According to \cite{39} $\tau_{n-n'}$ may be of the order of a few seconds if the influence of a hypothetical mirror magnetic field is taken into account.

We came to the conclusion  that the time evolution of the ``polarization vector'' $R_z(t)$ in quasi-free experiments is described at all times by the equation (\ref{27}). With $\beta$-decay included it may be rewritten in the following form \be R_z(t) = e^{-\left(\frac{1}{2}\lambda + \Gamma_{\beta}\right)\, t}\left( \cosh \Omega t + \frac{\lambda}{4 \Omega} \sinh \Omega t \right). \label{53x}\ee According to (\ref{53x}) the polarization vector $R_z$ shrinks in length without oscillations \cite{17}. It will never turn from $R_z=1,\,\rho_{11}=1,\,\rho_{22}=0$ to $R_z=-1,\,\rho_{11}=0,\,\rho_{22}=1$, i.e., from $\left| n \right>$ to $\left| \bar{n} \right>$ state. This may be viewed as Turing-Xeno-Watched Pot effect \cite{17,40}. For $\rho_{22}(t)=\left| \Psi_{\bar{n}}(t) \right|^2$ (\ref{53x}) yields \be \left| \Psi_{\bar{n}}(t) \right|^2 = \frac{4 \varepsilon^2}{\Omega^2}\,e^{-\left(\frac{1}{2}\lambda + \Gamma_{\beta}\right)\, t}\,\sinh^2\frac{\Omega\,t}{2}.\label{54x}\ee Equation (\ref{54x}) replaces the routine sine-squared expression for $\left| \Psi_{\bar{n}}(t) \right|^2$ (see (\ref{14})) which usually serves as a starting point in the discussions of the neutron to antineutron conversion. To retrieve the vacuum result (\ref{14}) with $d=0$ from (\ref{54x}) one simply puts $\lambda=0$.

\begin{figure}[htb]
\vspace{0.1cm}
\setlength{\unitlength}{1.0cm}
\begin{center}
\vspace{-0.1cm}
\begin{picture}(9.0,6.0)(0,0)
\put(0.35,0.35){\includegraphics[height=5cm]{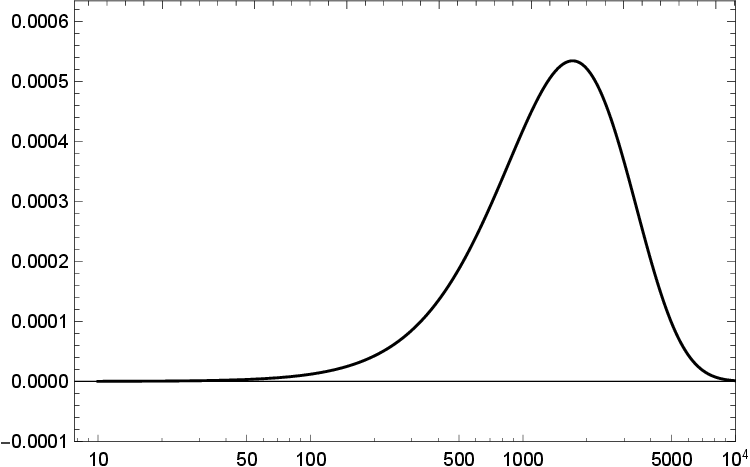}}
\put(0.5,5.5){\makebox(0,0)[cc]{$10^{7}\,\rho_{22}(t)$}}
\put(8.0,0.05){\makebox(0,0)[cc]{$t\,\,[\text{s}]$}}
\end{picture}
\vspace{-0.1cm}
\caption{The time dependence of $\left| \Psi_{\bar{n}}(t) \right|^2$ for the ILL experiment.}
\label{fig:01}
\end{center}
\end{figure}

The time dependence of $\left| \Psi_{\bar{n}}(t) \right|^2$ is plotted in Fig.\ref{fig:01} for the set of parameters corresponding to the ILL experiment with $\beta$-decay discarded: $\varepsilon = 1.16 \cdot 10^{-8}$ s$^{-1}$, $\lambda=0.37 \cdot 10^{-4}$ s$^{-1}$, $\Gamma_{\beta}=0$.

\smallskip

According to (\ref{54x}) for long times $t \gtrsim \lambda^{-1} \simeq 10^4$ s $\left| \Psi_{\bar{n}}(t) \right|^2$ displays an exponential decay due to the quantum damping (recall that we put $\Gamma_{\beta}=0$) \be \left| \Psi_{\bar{n}}(t) \right|^2 \simeq \frac{4 \varepsilon^2}{\lambda^2}\,\exp\left(-\frac{4\varepsilon^2}{\lambda}\,t\right), \label{55x} \ee with ${4 \varepsilon^2}/{\lambda^2} \simeq 4 \cdot 10^{-7}$, ${4 \varepsilon^2}/{\lambda} \simeq 10^{-11}$ s$^{-1}$. At short times $t \ll \lambda^{-1} \simeq 10^4$ s the asymptotics of (\ref{54x}) reads \be \left| \Psi_{\bar{n}}(t) \right|^2 \simeq \varepsilon^2 t^2 - \frac12 \varepsilon^2 \lambda t^3 + \ldots \label{56x} \ee The corresponding expansion of the Bloch vector component $R_z(t)$ is \be R_z(t) = 1 - 2 \varepsilon^2 t^2 + \ldots. \label{57x}\ee Deviation from $1$ starts quadratically and therefore repeated measurements will bring $R_z(t)$ back to one \cite{17,41}. Suppose that during the time interval $[0,\,T]$ $N$ ``measurements'' (annihilations) occur. Then at $N \to \infty$ \be R_z(t)\equiv \left[ 1 - 2 \varepsilon^2 \left( \frac{T}{N} \right)^2 \right]^N \to 1. \label{58x}\ee This means that the system remains ``frozen'' in the neutron state $\left|n\right>$.

\section{Conclusions and Outlook \label{results}}

In this paper a new approach to the analysis of the neutron-antineutron oscillations is proposed. It is based on the reduced density matrix formalism which describes the time evolution of the two-state system in contact with the environment. The role of the environment is correspondingly taken by the trap walls, the nuclear matter, and the residual gas. The contact with the surroundings results in the destruction of the off-diagonal elements of the density matrix and consequently the loss of the coherence. Or put another way, the departing information leads to the density matrix collapse.

The evolution equation for the Bloch vector of the $n-\bar{n}$ system is derived from the Lindblad equation. Under certain simplifying assumptions it is an equation for a damped oscillator. The damping parameter gives the rate at which the coherence is destroyed by the environment. For all three strategies to search for $n-\bar{n}$ oscillations the damping parameter is much larger than the mixing parameter. Therefore instead of the oscillatory dynamics the system displays non-oscillatory overdamped evolution described by (\ref{53x}) and (\ref{54x}). To perform a detailed analysis of the future ESS experiment \cite{35} one has to solve numerically the system of coupled Bloch equations (\ref{43x}) with a realistic set of the interaction parameters. In terms of the above classification the solution will be of the overdamped type.

Finally, a few words on the problems left for the future. It will be interesting to investigate the situation when the Lindblad equation can be reduced to Langevin equation describing the Brownian motion.  


 
\newpage

\vspace{1mm}
\begin{center}
{\sc{acknowledgments}}
\end{center}
\vspace{0.75mm}

This work has been supported by the Russian Science Foundation grant number 16-12-10414. The author thanks Yu.A.~Kamyshkov and L.J.~Varriano for collaboration during the initial stage of this project. I am indebted to M.I.~Krivoruchenko for discussions and to M.S.~Lukashov for comments at all stages of the work. The author thanks the participants of the INT-17-69W workshop ``Neutron-Antineutron Oscillations'' for many helpful remarks.








\end{document}